\documentclass[reprint, superscriptaddress, twocolumn, amsmath, amssymb, aps, prb]{revtex4-2}
\usepackage{graphicx}
\usepackage{dcolumn}
\usepackage{bm}
\usepackage{placeins}
\usepackage{appendix}
\usepackage{upgreek}

\usepackage{verbatim}
\usepackage[usenames,dvipsnames]{xcolor}
 \usepackage{amsmath}
 \usepackage{amsfonts}
 \usepackage{amssymb}
 \usepackage[colorlinks=true,citecolor=blue,linkcolor=red]{hyperref}

 \usepackage{bbold}     
 \usepackage[makeroom]{cancel}  
 \usepackage{multirow}    
 \usepackage[normalem]{ulem}        
 \usepackage{array}
 \usepackage{pdfpages}
\makeatletter
 \AtBeginDocument{\let\LS@rot\@undefined}
 \makeatother

\begin{document}


\title{Growth and fabrication of advanced PbTe-Pb quantum devices}

\author{Zehao Yu}
\email{equal contribution}
\affiliation{College of Semiconductors, Southern University of Science and Technology, Shenzhen 518055, China}
\affiliation{State Key Laboratory of Low Dimensional Quantum Physics, Department of Physics, Tsinghua University, Beijing 100084, China}

\author{Wenyu Song}
\email{equal contribution}
\affiliation{State Key Laboratory of Low Dimensional Quantum Physics, Department of Physics, Tsinghua University, Beijing 100084, China}

\author{Lining Yang}
\email{equal contribution}
\affiliation{State Key Laboratory of Low Dimensional Quantum Physics, Department of Physics, Tsinghua University, Beijing 100084, China}

\author{Zeyu Yan}
\email{equal contribution}
\affiliation{State Key Laboratory of Low Dimensional Quantum Physics, Department of Physics, Tsinghua University, Beijing 100084, China}

\author{Yuhao Wang}
\affiliation{State Key Laboratory of Low Dimensional Quantum Physics, Department of Physics, Tsinghua University, Beijing 100084, China}

\author{Weizhao Wang}
\affiliation{State Key Laboratory of Low Dimensional Quantum Physics, Department of Physics, Tsinghua University, Beijing 100084, China}

\author{Xinchen He}
\affiliation{State Key Laboratory of Low Dimensional Quantum Physics, Department of Physics, Tsinghua University, Beijing 100084, China}

\author{Ruidong Li}
\affiliation{State Key Laboratory of Low Dimensional Quantum Physics, Department of Physics, Tsinghua University, Beijing 100084, China}

\author{Jiaye Xu}
\affiliation{State Key Laboratory of Low Dimensional Quantum Physics, Department of Physics, Tsinghua University, Beijing 100084, China}

\author{Shuai Yang}
\affiliation{College of Semiconductors, Southern University of Science and Technology, Shenzhen 518055, China}

\author{Zonglin Li}
\affiliation{State Key Laboratory of Low Dimensional Quantum Physics, Department of Physics, Tsinghua University, Beijing 100084, China}

\author{Shan Zhang}
\affiliation{State Key Laboratory of Low Dimensional Quantum Physics, Department of Physics, Tsinghua University, Beijing 100084, China}

\author{Xiao Feng}
\affiliation{State Key Laboratory of Low Dimensional Quantum Physics, Department of Physics, Tsinghua University, Beijing 100084, China}
\affiliation{Beijing Academy of Quantum Information Sciences, Beijing 100193, China}
\affiliation{Frontier Science Center for Quantum Information, Beijing 100084, China}
\affiliation{Hefei National Laboratory, Hefei 230088, China}

\author{Tiantian Wang}
\affiliation{Beijing Academy of Quantum Information Sciences, Beijing 100193, China}
\affiliation{Hefei National Laboratory, Hefei 230088, China}

\author{Yunyi Zang}
\affiliation{Beijing Academy of Quantum Information Sciences, Beijing 100193, China}
\affiliation{Hefei National Laboratory, Hefei 230088, China}

\author{Lin Li}
\affiliation{Beijing Academy of Quantum Information Sciences, Beijing 100193, China}

\author{Runan Shang}
\affiliation{Beijing Academy of Quantum Information Sciences, Beijing 100193, China}
\affiliation{Hefei National Laboratory, Hefei 230088, China}

\author{Qi-Kun Xue}
\affiliation{State Key Laboratory of Low Dimensional Quantum Physics, Department of Physics, Tsinghua University, Beijing 100084, China}
\affiliation{Beijing Academy of Quantum Information Sciences, Beijing 100193, China}
\affiliation{Frontier Science Center for Quantum Information, Beijing 100084, China}
\affiliation{Hefei National Laboratory, Hefei 230088, China}
\affiliation{Department of Physics, Southern University of Science and Technology, Shenzhen 518055, China}

\author{Ke He}
\email{kehe@tsinghua.edu.cn}
\affiliation{State Key Laboratory of Low Dimensional Quantum Physics, Department of Physics, Tsinghua University, Beijing 100084, China}
\affiliation{Beijing Academy of Quantum Information Sciences, Beijing 100193, China}
\affiliation{Frontier Science Center for Quantum Information, Beijing 100084, China}
\affiliation{Hefei National Laboratory, Hefei 230088, China}
\affiliation{Department of Physics, Southern University of Science and Technology, Shenzhen 518055, China}

\author{Hao Zhang}
\email{zhanghao@sustech.edu.cn}
\affiliation{College of Semiconductors, Southern University of Science and Technology, Shenzhen 518055, China}


\begin{abstract}

We report recent progress in the growth and fabrication of advanced PbTe-Pb Majorana devices for three-terminal correlation experiments. A critical prerequisite for these devices is a continuous Pb film spanning from the PbTe nanowire to the substrate. In previous studies, PbTe nanowires were spatially separated from the substrate mask, making them unable to fulfill this requirement. To address this issue, we have developed a new growth and fabrication approach. Key fabrication steps are introduced between the PbTe growth and Pb deposition. The top facet of the PbTe nanowire is covered with a Te capping layer before breaking the ultra-high vacuum, enabling \textit{ex situ} processing for subsequent steps. The mask is then removed, followed by the fabrication of shadow walls. After reloading the chip into the growth chamber, the Te protection layer is removed through thermal decomposition, and the Pb film is subsequently deposited. Notably, a pristine PbTe-Pb interface with an \textit{in-situ}-equivalent quality can be achieved despite the \textit{ex-situ} processing. Our technique facilitates the develop of advanced PbTe-Pb quantum devices with complex designs, paving the way toward the realization of topological qubits.

\end{abstract}

\maketitle  

\section{Introduction}

Selective-area-grown (SAG) PbTe nanowires \cite{Jiangyuying, Erik_PbTe_SAG, CaoZhanPbTe} have recently emerged as a promising material platform for the realization of Majorana zero modes and topological quantum computation \cite{ReadGreen, Kitaev, Lutchyn2010, Oreg2010, RMP_TQC}. Built upon the rapid progress in its growth and transport characterizations \cite{PbTe_AB, Fabrizio_PbTe, Zitong_JJ, Wenyu_QPC, Yichun_gap, Yuhao_QPC, Ruidong_PJJ, Vlad_PbTe, PbTe_In, Yichun_SQUID, Yuhao_degeneracy, Wenyu_Disorder, Quantized_Andreev, Zonglin_Anisotropy}, robust zero modes can now be routinely observed in PbTe-Pb hybrid nanowires \cite{Zhangshan_ZBP, Ruidong_ZBP}. While these observations are consistent with a Majorana interpretation, alternative trivial mechanisms, such as quasi-Majoranas \cite{TudorQuasi, WimmerQuasi}, cannot be ruled out due to the limitations of the current device configuration, i.e., local tunneling spectroscopy. To search for stronger evidence, more complex device configurations, such as three-terminal devices and topological qubits, are desired \cite{Andreev_rectifier, 2010_Alicea, 2017_Box_qubit, 2017_PRB_Scalable, MS_Blueprint}.

A key requirement for these advanced devices is a superconducting film connecting multiple nanowires. Crucially, the film needs to span continuously from the nanowire to the substrate. This seemingly-straight-forward design is, however, non-trivial for PbTe nanowires. In recently developed high-quality SAG devices \cite{Wenyu_Disorder, Quantized_Andreev, Yuhao_degeneracy}, the PbTe nanowires were designed to sit below and well separated from the Si$_3$N$_4$ mask. This design eliminates disorder originating from the mask, enables self-organized growth, and results in high-quality ballistic transport. Despite this significant advantage, a notable drawback is that the Pb film on the wire is disconnected from the Pb film on the substrate. As previously noted, a continuous superconducting film spanning from the nanowire to the substrate is required for three-terminal experiments and the realization of topological qubits \cite{MS_2023, NextSteps, Prada2020, Leo_perspective, MS_Blueprint}. Filling this gap and bridging the disconnected regions turn out to be a challenging task. Growing thicker PbTe to align with the mask could solve this issue; however, it significantly increases device disorder due to direct contact with the mask.

To realize a continuous superconducting film while still maintaining a low disorder level, we have developed a new technical approach. After the PbTe growth using molecular beam epitaxy (MBE), we covered the nanowires \textit{in situ} with a thin layer of Te. The chip was then taken out and the mask was removed using ultrasonic cleaning. After the fabrication of shadow walls, we reloaded the chip into the MBE chamber. The Te protection layer was removed via a short Ar treatment and thermal annealing, resulting in a pristine PbTe surface for Pb deposition. This approach enables the realization of complex PbTe-Pb quantum devices while maintaining their high material quality. In the following sections, we present and characterize the key fabrication steps.

\section{Mask removal}

\begin{figure*}[htb]
\includegraphics[width=\textwidth]{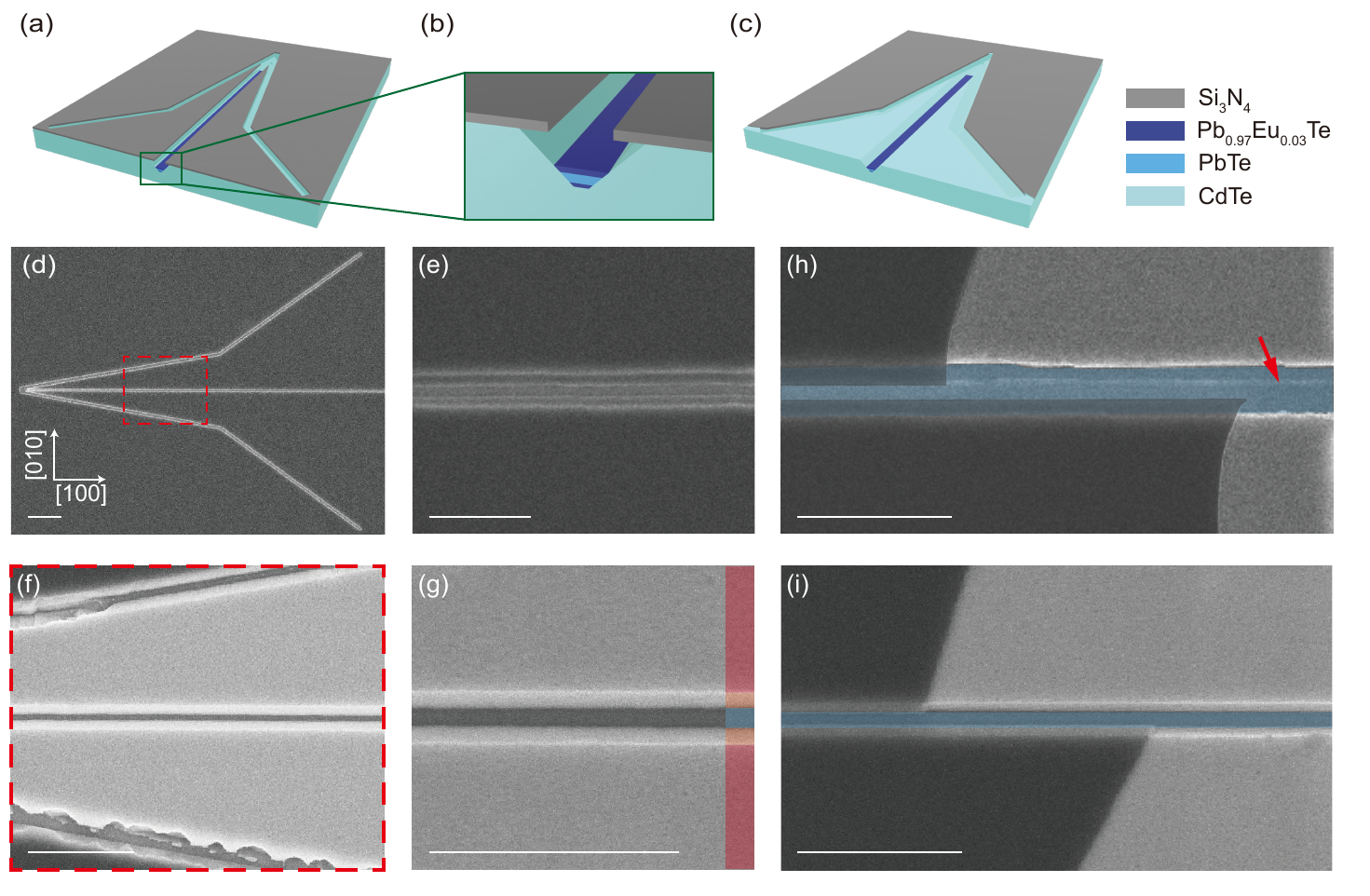}
\centering
\caption{Mask removal. (a-c) Schematics of the trench pattern. The Si$_3$N$_4$ mask in (a-b) is peeled off in (c) using ultrasonic cleaning. (d) SEM of the mask pattern. (e) SEM of a PbTe nanowire before the mask removal. (f) SEM after the mask removal. The region in (f) corresponds to the red dashed box in (d). (g) Magnified SEM focusing on the middle wire. (h-i) SEMs of two examples where part of the mask was removed. Scale bars in (d-i), 1 $\upmu$m.  }
\label{fig1}
\end{figure*}

Figures 1(a-c) show the essence of mask removal. A CdTe(001) substrate was first covered with a 10-nm-thick of Si$_3$N$_4$ layer deposited via plasma enhanced chemical vapor deposition. Nanowire patterns were defined and etched using reactive ion etching. Different from our previous patterns, we introduced two auxiliary segmented lines on each side of the main nanowire trench. As shown in Fig. 1(a), these auxiliary lines converge with one end of the nanowire trench. This mask pattern is specially designed for the subsequent lift-off of the mask. The substrate was then loaded into an MBE chamber. An initial Ar ion milling was performed to remove CdTe surface oxides, followed by thermal annealing and the SAG of a Pb$_{0.97}$Eu$_{0.03}$Te buffer layer, the PbTe nanowire, and a Pb$_{0.97}$Eu$_{0.03}$Te capping layer. The growth details are identical to those reported in our previous work \cite{Wenyu_Disorder}. 

\begin{figure*}[htb]
\includegraphics[width=\textwidth]{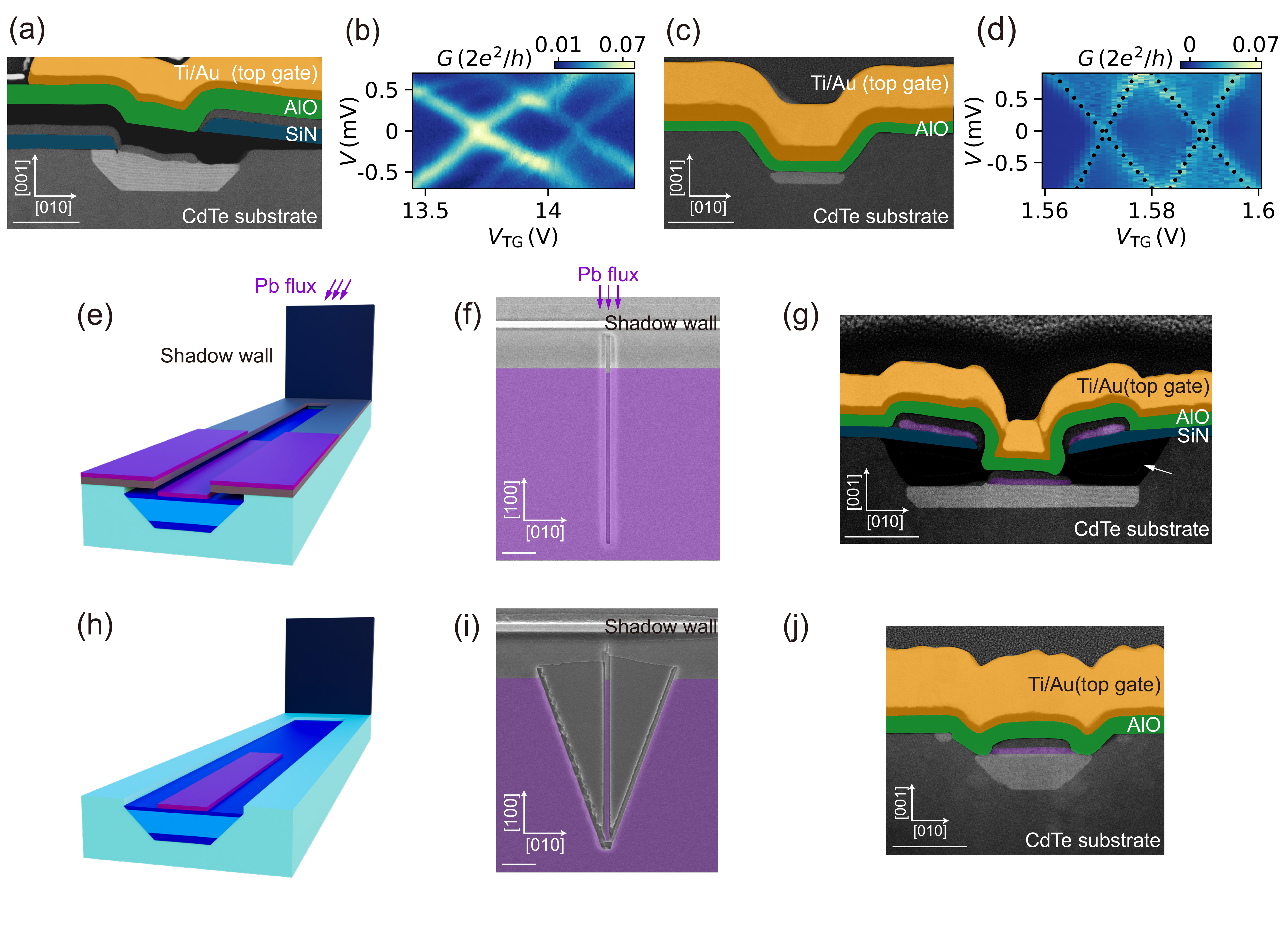}
\centering
\caption{Gate tunability before and after mask removal. (a) False-colored cross-sectional STEM of a PbTe QD with the presence of the Si$_3$N$_4$ mask (labeled as SiN). The partial detachment of the right mask (blue) from the substrate is caused by focused ion beam during the STEM process.  (b) Stability diagram of the device in (a). (c) Cross-sectional STEM of a PbTe QD in the absence of the Si$_3$N$_4$ mask. (d) Stability diagram of the device in (c). (e-g) Schematic, SEM and STEM of a PbTe-Pb nanowire in the presence of the mask. (h-j) Similar to (e-g) but with the mask being removed. Scale bars are are 100 nm in (a), (c), (g), (j), and 1 $\upmu$m in (f) and (i).     }
\label{fig1}
\end{figure*}

Figure 1(b) presents a magnified cross-sectional schematic of the nanowire-mask geometry, highlighting that the PbTe nanowire is well separated from the Si$_3$N$_4$ mask. The mask edges become suspended due to the thermal decomposition of the underneath CdTe substrate during annealing. This undercut alone is insufficient to peel off the mask. The presence of the two outer broken lines can significantly enhance this possibility, especially at the terminus of the middle wire where these lines intersect. Most of the mask near this intersection is fully suspended due to the extensive decomposition of the underlying CdTe. The substrate was then immersed into an acetone bath and subjected to ultrasound agitation. The ultrasonic vibration peels off the mask region delineated by the two broken lines, starting from the intersection point. Figure 1(c) shows the schematic geometry after mask removal. The exposed middle wire can then be used for subsequent fabrications.

To demonstrate its feasibility, we show the scanning electron micrographs (SEMs) of each key step. Figure 1(d) displays the trench pattern. After nanowire growth and before mask removal, the SEM of the middle wire is shown in Fig. 1(e). The mask edges block the view of the nanowire regions located within the undercut region. After applying ultrasound (Fig. 1(f)), the mask in the middle of the pattern is successfully peeled off (the bright region). Figure 1(g) is a magnified view of the exposed middle wire. The PbTe nanowire (false-colored blue at the right region) and its edges are clearly identified. The neighboring CdTe facets (tilted slopes, false-colored orange) and their transition to the flat CdTe substrate (false-colored red) are also clearly visible without the mask coverage. 

This unobstructed visibility allows for the pre-selection of high-quality nanowires with uniform surfaces and edges for subsequent fabrication, increasing the overall device yield. In our previous work, the wire edges were blurred due to the mask coverage (Fig. 1(e)), making it difficult to assess the wire quality prior to fabrication. Figure 1(h) shows a direct comparison on a single wire where only a part of the mask is removed while the other part remains. Based on the right region without mask coverage, we can observe that the wire edges are rough and the surface is non-uniform. The red arrow marks a small bump on the PbTe surface. In the Appendix Fig. 6, we show a scanning transmission electron microscopy (STEM) characterization of such surface bumps. These critical structural details, however, remain hidden in the left part of Fig. 1(h) due to the mask coverage. Consequently, without the mask removal step, low-quality wires cannot be effectively filtered out prior to fabrication. For comparison, Fig. 1(i) shows the case of a verified high-quality nanowire both with and without the presence of the mask. 

\begin{figure*}[htb]
\includegraphics[width=\textwidth]{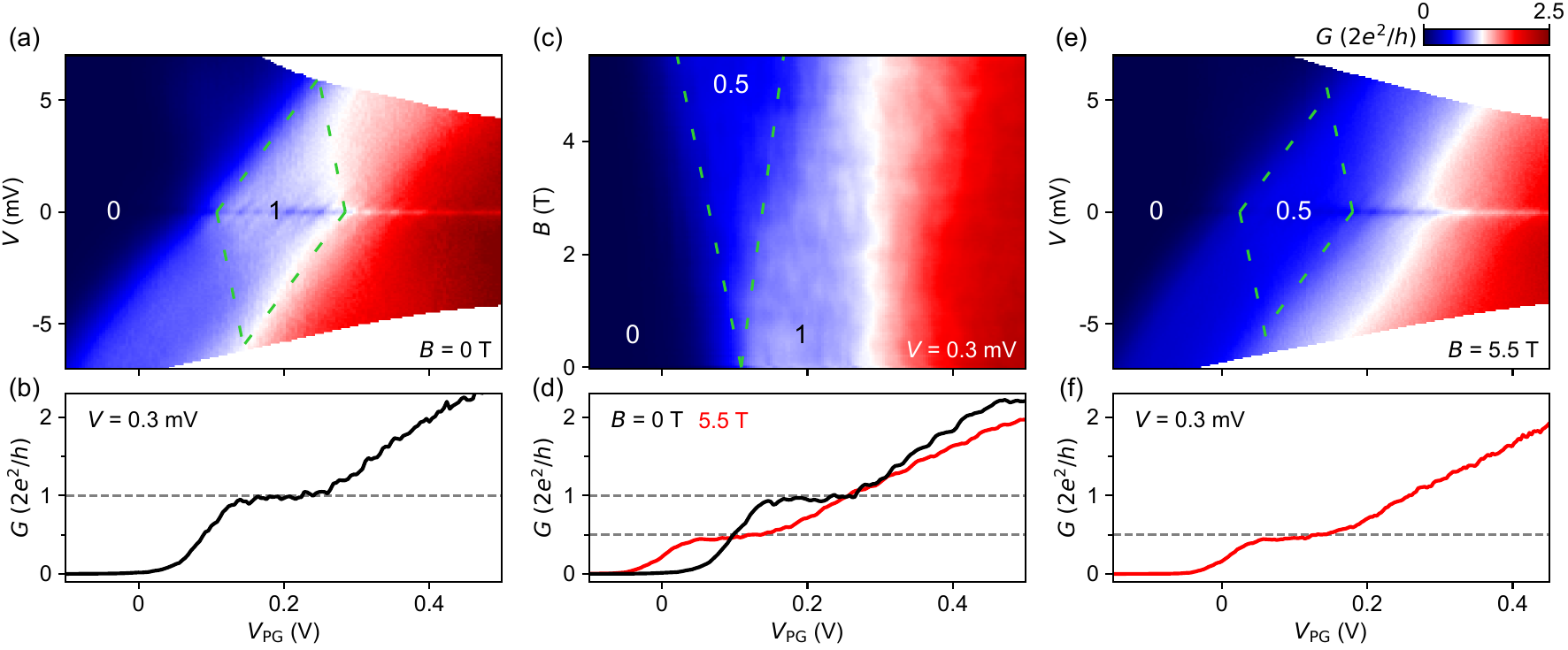}
\centering
\caption{Ballistic transport after mask removal. (a) $G$ vs $V$ and $V_{\text{PG}}$ at 0 T. (b) Line cut from (a) near zero bias ($V$ = 0.3 mV). (c) $B$ scan at $V$ = 0.3 mV. The $B$ direction is out-of-plane. (d) Line cuts at 0 T(black) and 5.5 T (red), respectively. (e) $G$ vs $V$ and $V_{\text{PG}}$ at 5.5 T. (a-e) share the same color bar. (f) Line cut from (e) near zero bias ($V$ = 0.3 mV). The measurement was two-terminal. A series resistance, including the fridge filters and a contact resistance of 3 k$\Omega$, has been subtracted. }
\label{fig1}
\end{figure*}

An additional advantage of mask removal is the significantly enhanced gate tunability, a critical factor for device control. We quantify this tunability using the gate lever arm, $\alpha$, which measures the amount of energy that can be tuned by 1 V of gate voltage. Figure 2(a) presents a cross-sectional STEM of a typical device with the presence of mask. For its SEM, we refer to Fig. 7 in the Appendix. This device exhibits quantum dot (QD) behavior, as shown in Fig. 2(b). From the diamonds in the charge stability diagram, we can extract $\alpha \sim$ 1.1 meV/V, i.e. a 1 V variation in the gate voltage ($V_{\text{TG}}$) shifts the dot energy (electro-chemical potential) by 1.1 meV. For comparison, Fig. 2(c) displays an STEM of a device fabricated following mask removal. The lever arm extracted from the QD diamond (outlined by dashed lines in Fig. 2(d)) is $\sim$ 60 meV/V, which is significant larger than that obtained in Fig. 2(b). Although the reduction in Al$_2$O$_3$ gate dielectric thickness (30 nm in Fig. 2(a) vs 15 nm in Fig. 2(c)) contributes to this increase, the mask removal also plays a pivotal role. Without the mask, the Al$_2$O$_3$ dielectric layer (false-colored green) and the top gate (orange) can be conformally attached onto the nanowire surface. Conversely, in Fig. 2(a), the physical separation introduced by the mask between the gate and the nanowire severely diminishes its tunability. Furthermore, the Al$_2$O$_3$ layer also needs to be sufficiently thick to ensure its continuous step coverage across the mask edges. To further quantify the enhanced gate tunability, Fig. 7 in the Appendix shows the simulated gate capacitance for the same wire with and without mask removal.

The gate issue (in the presence of the mask) becomes more detrimental for superconducting devices, as shown in Figs. 2(e-j). Figures 2(e) and 2(f) present a schematic and an SEM, respectively, of a PbTe-Pb device in the presence of the mask. The shadow wall selectively mask part of the wire from Pb deposition, defining regions for normal contacts and tunnel barriers. The superconducting Pb film is false-colored violet. Figure 2(g) displays a cross-sectional STEM acquired after device fabrication. The top gate (orange) is designed to tune the electro-chemical potential in the proximitized PbTe region. However, the presence of the Pb film on both the PbTe nanowire (middle) and atop the adjacent mask (left and right) induces strong electrostatic screening, suppressing this tunability. Moreover, the lateral sides of the PbTe nanowire are spatially displaced further away from the top gate due to mask undercut (see the white arrow). Together, these factors strongly limit the gate tunability. The protocol of mask removal effectively solves this issue, as depicted in Figs. 2(h-j). Figure 2(i) shows that the parasitic Pb film previously resting on the mask is removed, mitigating its screening over the lateral regions of the PbTe nanowire. The mask undercut also disappears, and the gate can be positioned in closer proximity to the wire surface. For PbTe-Pb devices fabricated following this approach, better control of zero-energy modes can be achieved and will be reported in a separate work.

\begin{figure*}[htb]
\includegraphics[width=\textwidth]{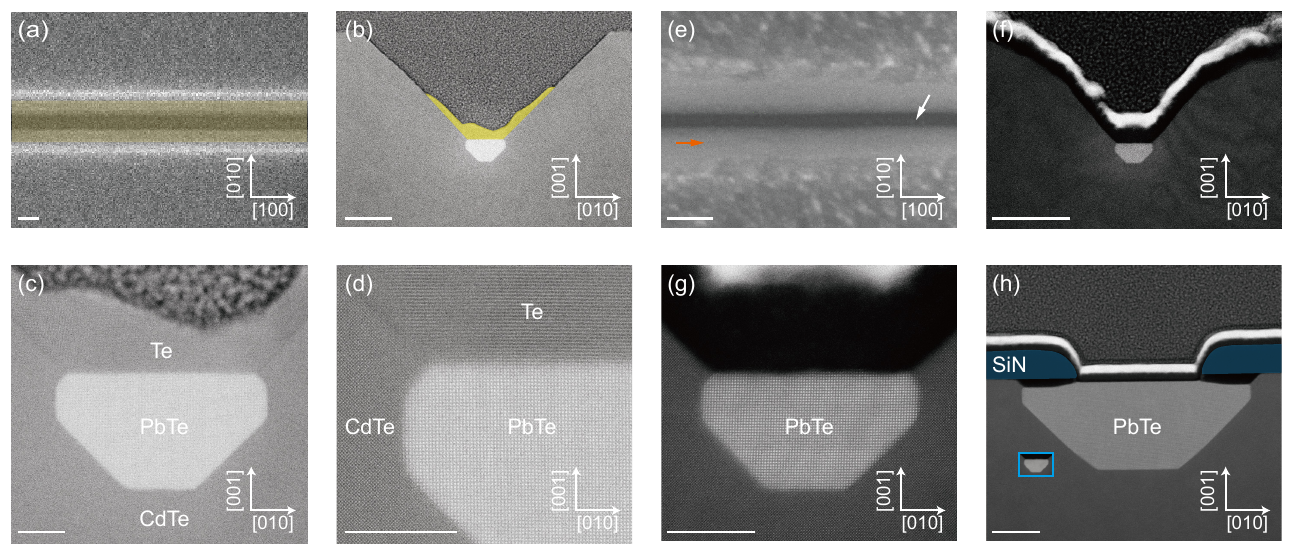}
\centering
\caption{Te capping and its removal on PbTe nanowires. (a) SEM of a PbTe nanowire with Te capping after the mask removal. The Te layer is false-colored yellow in (a-b). Scale bar is 100 nm. (b-d) Cross-sectional STEMs. Scale bars are 50 nm for (b) and 10 nm for (c-d). The Pb$_{0.97}$Eu$_{0.03}$Te buffer and capping layers are indistinguishable from PbTe, thus all labeled as PbTe. (e) SEM after the Te removal. Scale bar, 100 nm. (f-g) Cross-sectional STEMs of the wire in (e). Scale bars are 50 nm for (f) and 10 nm for (g). The dark region above PbTe in (f-g) is evaporated Ti before the FIB process. (h) Comparison of the wire in (g) (shown as an inset) with the wire in our previous work. Scale bar is 50 nm.  }
\label{fig1}
\end{figure*}

To verify whether the mask removal deteriorates the device quality, Figure 3 shows the transport results of a representative device fabricated using this protocol \cite{note_previous}. The PbTe nanowire was grown following the steps in Figs. 1(a-c), after which contacts and top gates were fabricated. The Al$_2$O$_3$ dielectric layer was grown using atomic layer deposition. The PbTe channel length, i.e., spacing between contacts, is around 450 nm, and the wire width is around 95 nm. 

Figure 3(a) maps the differential conductance ($G\equiv dI/dV$) as a function of bias voltage ($V$) and gate voltage, measured in a dilution fridge at a base temperature below 50 mK. This device incorporates three top gates (in a QD configuration). We swept the central gate ($V_{\text{PG}}$) and the two side gates ($V_{\text{TG}}$) simultaneously while maintaining a constant offset ($V_{\text{PG}}-V_{\text{TG}}$ = 0.25 V). The dashed lines mark a diamond shape, indicating the presence of a ballistic plateau. The minor dip at zero bias is likely due to dissipative tunneling \cite{ZhangShan_dissipation, WangZhichuan_weak_dissipation}. This dip feature can be routinely observed in PbTe nanowires with small cross-sectional areas \cite{Yuhao_dot}, but is absent in those with large areas \cite{Yuhao_degeneracy, Wenyu_Disorder}, indicating different electro-magnetic environments. 

Figure 3(b) shows a line cut at $V$ = 0.3 mV, i.e. near zero bias but outside the small dip. A contact resistance of 3 k$\Omega$ together with fridge filters have been subtracted from this two-terminal circuit. A plateau is clearly resolved near the quantized value of $2e^2/h$. Based on the diamond shape in Fig. 3(a) (labeled as ``1''), we estimate the subband spacing to be $\sim$ 6 meV and the gate lever arm to be 34 meV/V. For this device, the Al$_2$O$_3$ dielectric gate layer beneath the PG is $\sim$ 30 nm thick. This value is consistent with the lever arm extracted from the QD spectroscopy in Fig. 2.  As a comparison, the lever arms in PbTe quantum-point-contact devices with a side-gate configuration are much smaller, generally below 7 meV/V \cite{Yuhao_degeneracy, Wenyu_Disorder}. Improved gate control with larger lever arms in quantum devices is crucial for Majorana research.

Applying a magnetic field ($B$) induces the Zeeman splitting of the first subband, as shown in Fig. 3(c) by the emergence of a $e^2/h$ plateau (labeled as ``0.5'' between the dashed lines). Here, $e$ is the electron charge and $h$ is the Planck constant.   The red curve in Fig. 3(d) is the line cut at $B$ = 5.5 T. We also performed the bias spectroscopy on this $e^2/h$ plateau (Figs. 3(e-f)). From the diamond size of $\sim$5.3 meV, an effective Land\'e g-factor of $\sim$16 can be extracted. The ballistic transport, the subband spacing, and the extracted g-factor are consistent with those reported in our previous work for devices retaining the mask \cite{Yuhao_QPC, Yuhao_degeneracy}. These results confirm that the pristine, low-disorder transport channel is well-preserved during the new fabrication steps, i.e., the mask removal. Meanwhile, the gate lever arm is significantly enhanced.

\section{T\MakeLowercase{e} protection of P\MakeLowercase{b}T\MakeLowercase{e} nanowires}

Following mask lift-off, a continuous Pb film can be deposited, bridging the PbTe nanowire and the substrate. However, breaking the ultra-high vacuum during the removal process inevitably induces surface oxidation. This oxide layer must be strictly avoided, as it introduces disorder and severely degrades the PbTe-Pb interface \cite{Takei2013}. To address this issue, the nanowire surface needs to be protected prior to vacuum breaking. Moreover, this protective layer must be removed thoroughly and non-destructively after the vacuum breaking and before the subsequent MBE growth of the Pb film. We choose Te as the protection layer. After the PbTe growth, a thin layer of Te was grown \textit{in situ} before breaking the vacuum. Then the dielectric mask and the Te layer on top of it were removed using ultrasound agitation. The Te coating on the nanowire remains intact. Figure 4(a) shows the SEM after this process. The cross-sectional STEMs are shown in Figs. 4(b-d). The Te capping layer seamlessly adheres to the PbTe top facet (even under atomic-level resolution), protecting it from oxidation and environmental damage.  

\begin{figure*}[htb]
\includegraphics[width=\textwidth]{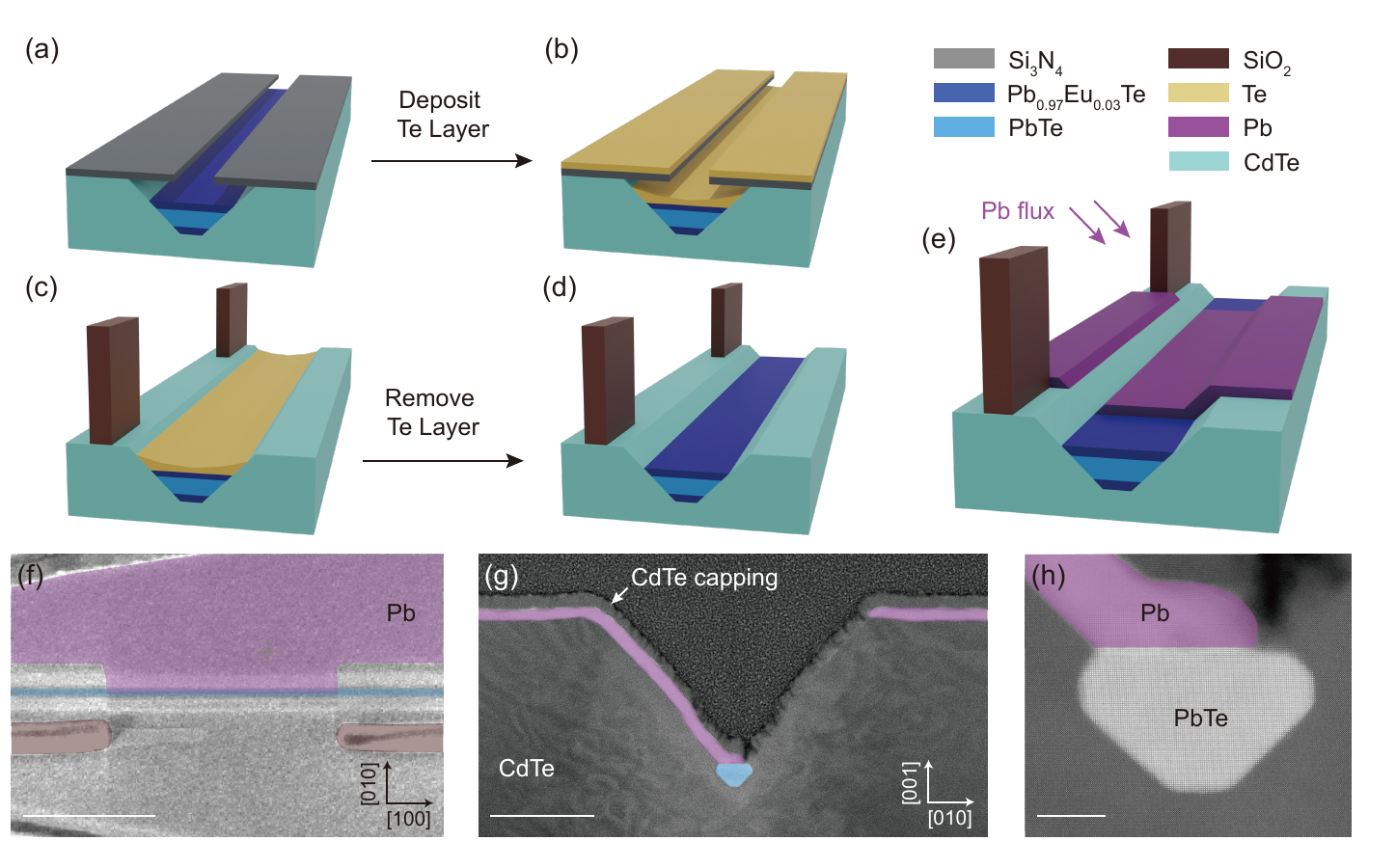}
\centering
\caption{Realization of three-terminal PbTe-Pb devices. (a) Schematic of the nanowire after the PbTe/Pb$_{0.97}$Eu$_{0.03}$Te growth. (b) \textit{In situ} deposition of the Te layer. (c) Removal of Si$_3$N$_4$ dielectric mask and fabrication of shadow walls. (d) Decomposition of the Te layer. (e) Deposition of Pb film and CdTe capping (not shown). (f) SEM of a PbTe-Pb three-terminal nanowire after the step of (e). Scale bar, 1 $\upmu$m. The Pb film is false colored violet and PbTe is colored blue. (g-h) STEMs of the cross section of (f). Scale bars are 100 nm for (g) and 10 nm for (h). The white arrow marks the CdTe capping layer above the Pb film. }
\label{fig1}
\end{figure*}

We next demonstrate that this Te layer can be thoroughly and cleanly removed. The chip was reloaded into the MBE chamber. A short Ar ion milling was performed, followed by thermal annealing at 340 $^{\circ}$C for 60 minutes. This temperature regime is sufficient to decompose and desorb the Te layer completely while the underlying PbTe nanowire remains structurally intact. Figure 4(e) shows the SEM after this process. The Te capping layer is entirely eliminated, leaving behind a pristine PbTe top surface (indicated by the white arrow). The adjacent CdTe facets (marked by the orange arrow) also become clean, further corroborating the complete absence of Te. Figures 4(f) and 4(g) show the cross-sectional STEM. To protect the STEM lamella during focused ion beam (FIB) milling, we evaporated a thin bilayer of Ti and Au (shown as the dark and white layers directly on top of the PbTe). Figure 4(g) reveals that, on an atomic-resolution scale, the PbTe top surface is intact and the Te capping layer is removed with no observable residue. This pristine surface provides an optimal template for the subsequent epitaxial deposition of the Pb film.    

We further note that the PbTe cross-section area is substantially reduced compared to those reported in our previous works \cite{Yuhao_degeneracy, Wenyu_Disorder}. For a direct comparison, Fig. 4(h) shows the cross section of a typical device from our previous works, while the inset is the device in Fig. 4(g) plotted under the same scale bar. This ultra-small cross section enables the realization of a finite charging energy in PbTe quantum dots (see Ref. \cite{Yuhao_dot} for a systematic study). More importantly, as demonstrated in Fig. 3, the ballistic transport remains despite the small area and the increased surface-to-volume ratio.

\section{Three-terminal P\MakeLowercase{b}T\MakeLowercase{e}-P\MakeLowercase{b} nanowires }

The combined capabilities of mask removal and Te protection allow us to realize three-terminal PbTe-Pb devices. Figures 5(a-e) show the schematics of the key fabrication steps. Figure 5(a) depicts a PbTe nanowire grown following our standard procedure \cite{Wenyu_Disorder}. The Pb$_{0.97}$Eu$_{0.03}$Te bottom buffer layer (dark blue) ensures the flatness of the bottom facet, while the top Pb$_{0.97}$Eu$_{0.03}$Te layer modulates the coupling between the PbTe and Pb. Immediately after this growth, we deposited a think capping layer of Te \textit{in situ}, as shown in Fig. 5(b). The chip was sequentially tilted toward one side and then the other so that the impinging Te flux could cover both sides of the CdTe facets underneath the mask. The chip was then taken out from the MBE chamber. After ultrasound cleaning for mask removal, we fabricated two shadow walls near the nanowire using Hydrogen SilsesQuioxane (HSQ) \cite{Jiangyuying}, as seen in Fig. 5(c). The shadowed regions define the areas for tunnel barriers and normal contacts.

After these processes, the chip was reloaded into the MBE chamber. A short Ar treatment followed by thermal annealing removes the Te layer (Fig. 5(d)). The chip was then cooled using liquid nitrogen, and a Pb film was deposited at a tilted angle. Without the mask, the Pb film can continuously extend from the PbTe/ Pb$_{0.97}$Eu$_{0.03}$Te wire surface to the CdTe slope, and finally onto the substrate surface where a macroscopic contact can be made. Finally, a thin layer of CdTe was grown (not shown) to cap the chip, preventing the Pb film from oxidization.

Figure 5(f) shows the SEM of a typical wire grown using this approach. The growth details can be found in the Appendix. For clarity, the nanowire is false-colored blue and the Pb film is colored violet. To confirm that the Pb film can continuously bridge the gap from the nanowire to the substrate, Fig. 5(g) shows the cross-sectional STEM. Both the Pb film and the CdTe capping (white arrow) are structurally continuous, solving the bottleneck for three-terminal devices. Figure 5(h) shows a high-resolution view of the PbTe-Pb interface. Despite the vacuum breaking during the \textit{ex-situ} transfers, the interface remains atomically sharp. This suggests that Te protection is robust and its removal is complete, ensuring a sharp PbTe-Pb interface with \textit{in-situ}-equivalent quality. We note that this method of using a protection-and-removal layer may also be applied to other semiconductor-superconductor material systems (such as InSb), especially for those where the semiconductor and superconductor cannot be incorporated into the same MBE system and where vacuum breaking is unavoidable.

\section{Summary}

To summarize, we have reported a new growth and fabrication approach for PbTe-Pb nanowires. This approach addresses the discontinuity issue that previous PbTe-Pb hybrid nanowires suffered from. Using a specialized mask design, the dielectric mask layer can be removed via ultrasonic cleaning. The mask removal greatly enhanced the gate tunability in quantum devices. Capping the PbTe nanowire with a Te layer prevents its damage and oxidation during the \textit{ex situ} processing. The complete thermal decomposition of Te ensures a clean and intact PbTe interface for subsequent Pb film deposition. This capping-and-removing technique lifts the restriction of uninterrupted \textit{in situ} vacuum for achieving a pristine superconductor-semiconductor interface, and may have profound implications for hybrid quantum devices beyond PbTe. Finally, a continuous Pb film spanning from the PbTe surface (with \textit{in-situ}-equivalent quality) to the device substrate can be reliably achieved. This approach opens the door to advanced Majorana devices, such as three-terminal end-to-end correlation experiments. Moreover, two spatially separated PbTe nanowires can also be connected by a Pb film using this method, enabling more sophisticated architectures such as topological qubits.

\section{Acknowledgment}

This work is supported by National Natural Science Foundation of China (92565302) and Quantum Science and Technology-National Science and Technology Major Project (2021ZD0302400). This work used the facilities of the National Center for Electron Microscopy in Beijing at Tsinghua University. W. S. acknowledges the Postdoctoral Fellowship Program and China Postdoctoral Science Foundation (Grant No. BX20250167 and No. 2025M783399). S.Y. acknowledges the China Postdoctoral Science Foundation (Grant No. 2024M751610) and Postdoctoral Fellowship Program of China Postdoctoral Science Foundation (Grant No. GZC20231368).

\section{Data availability}

The data that support the findings of this article are openly available at Ref. \cite{rawdata}.

\appendix

\section{STEM of a PbTe nanowire with a non-uniform surface}

\begin{figure}[htb]
\includegraphics[width=\columnwidth]{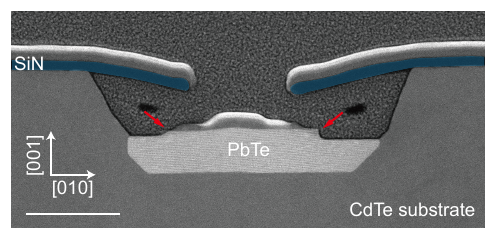}
\centering
\caption{Cross-sectional STEM of a PbTe nanowire with a rough surface. The two red arrows mark two bumps on the PbTe surface, indicating a low wire quality. These bumps cannot be clearly identified under SEM prior to fabrication due to the coverage of the Si$_3$N$_4$ mask (blue). Only after the mask removal can they be clearly visible via SEM (the red arrow in Fig. 2(h)). On top of the PbTe nanowire (and the mask), we capped a thin layer of CdTe (dark) and Ti/Au (bright). The scale bar is 100 nm.   }
\label{fig1}
\end{figure}

\section{SEMs of the two devices in Fig. 2}

\begin{figure}[htb]
\includegraphics[width=\columnwidth]{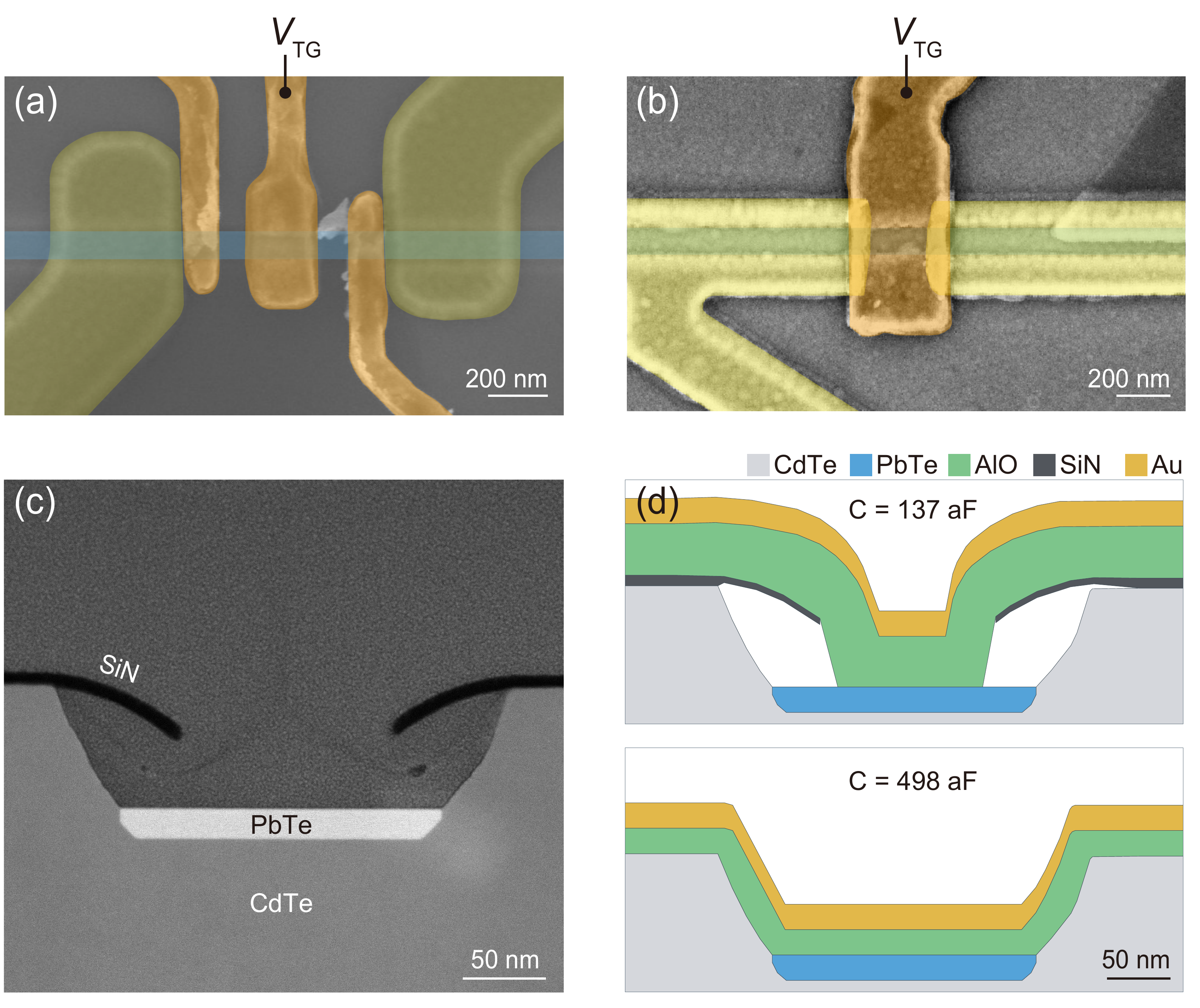}
\centering
\caption{(a) and (b): False-color SEMs of the devices corresponding to Figs. 2(a-b) and Figs. 2(c-d), respectively. The PbTe nanowires are shown in blue, the contacts in yellow, and the top gates in orange. The $V_{\text{TG}}$ gates in Figs. 2(b) and 2(d) (the x-axis) are marked in (a-b). The dielectric layer is Al$_2$O$_3$ grown by atomic layer deposition. (c) Cross-sectional STEM of a PbTe nanowire with a suspended mask. (d) Dielectric and gate layouts of this wire for two fabrication techniques: without (upper) and with (lower) mask removal. The gate-wire capacitance was simulated using a finite element method and is labeled in each panel: 137 aF and 498 aF, respectively, corresponding to a 3.6-fold enhancement in gate tunability. The contact spacing was assumed to be 500 nm. The Al$_2$O$_3$ dielectric layer is 40 nm thick in the upper panel to ensure continuity across the mask, whereas it is 20 nm thick in the lower panel. In practice, a thinner Al$_2$O$_3$ layer (e.g., 10 nm) is sufficient to provide continuous coverage after mask removal, and the tunability can be further enhanced.}
\end{figure}

\section{Details of device growth and fabrication}

\textbf{Dielectric trenches.} A 10-nm-thick Si$_3$N$_4$ mask was deposited on a CdTe(001) substrate using the plasma-enhanced chemical vapor deposition. Then AR-P 672.045 resist was spin-coated at 6000 rpm for 60 s and baked at 130 $^{\circ}$C for 5 min. Patterns of nanowire trenches (shown in Fig. 1) and markers were defined using electron beam lithography (EBL). The chip was then developed in a solution (MIBK: IPA (1:3)) for 40 s. Reactive ion etching was performed to etch the exposed region of the Si$_3$N$_4$, forming the dielectric mask. The resist was removed with acetone and then cleaned for 4 minutes in oxygen plasma.

\textbf{PbTe growth and Te capping.} The pre-patterned substrate was loaded into an MBE chamber. Ar ion bombardment (2.7 keV, 1.50 $\upmu$Acm$^{-2}$) was performed for 24 minutes to remove the CdTe surface oxides inside the trenches. To ensure a clearn CdTe surface, the chip was then annealed at 350 $^{\circ}$C for 80 min. We then set the temperature at 340 $^{\circ}$C for the following SAG growth: 5-nm-thick Pb$_{0.97}$Eu$_{0.03}$Te buffer, 15-nm-thick PbTe nanowire and 5-nm-thick Pb$_{0.97}$Eu$_{0.03}$Te top layer. Finally, the sample was capped by a 10-nm-thick Te layer \textit{in situ} at room temperature.

\textbf{Mask removal and shadow-wall fabrication.} The Si$_3$N$_4$ mask was removed via ultrasonication in acetone. Hydrogen SilsesQuioxane (FOx-16, Dow Corning) resist was spin-coated at 2000 rpm for 60 s and baked at 150 $^{\circ}$C for 150 s. Shadow walls and markers were defined by EBL and developed in tetramethylammonium hydroxide (TMAH) aqueous solution for 155 s. 

\textbf{Te capping removal.} The sample was re-loaded into the MBE chamber. Ar-ion bombardment (2 keV, 1.50 $\upmu$Acm$^{-2}$) was performed for 1 minute to remove the surface oxide of the Te capping layer. The chip was then annealed at 340 $^{\circ}$C for 60 min to decompose the Te capping layer.

\textbf{Pb deposition and capping.} The sample was transferred \textit{in situ} to another sample stage within the same MBE chamber and cooled to $\sim$100 K using liquid nitrogen. A 9-nm-thick Pb film was deposited at a tilted angle. Finally, the entire chip was capped with a 14-nm-thick CdTe layer.

\bibliography{mybibfile}




\end{document}